\documentclass[iop]{emulateapj}
\usepackage[usenames, dvipsnames]{color}
\usepackage[colorlinks=true,linkcolor=black,citecolor=blue,urlcolor=black]{hyperref}%

\slugcomment{Accepted for publication in The Astrophysical Journal: July 31, 2018}

\shorttitle{Solar Torque Variability}
\shortauthors{A. Finley, S. Matt \& V. See}

\begin{document}


\title{The Effect of Magnetic Variability on Stellar Angular Momentum Loss I: \\ The Solar Wind Torque During Sunspot Cycles 23 \& 24}


\author{Adam J. Finley*, Sean P. Matt \& Victor See}
\affil{University of Exeter,
              Devon, Exeter, EX4 4QL, UK}
\email{*af472@exeter.ac.uk}



\begin{abstract}
The rotational evolution of cool stars is governed by magnetised stellar winds which slow the stellar rotation during their main sequence lifetimes. Magnetic variability is commonly observed in Sun-like stars, and the changing strength and topology of the global field is expected to affect the torque exerted by the stellar wind. We present three different methods for computing the angular momentum loss in the solar wind. Two are based on MHD simulations from \cite{finley2018dipquadoct}, with one using the open flux measured in the solar wind, and the other using remotely-observed surface magnetograms. Both methods agree in the variation of the solar torque seen through the solar cycle and show a $30-40\%$ decrease from cycle 23 to 24. The two methods calculate different average values, $2.9\times 10^{30}$ erg (open flux) and $0.35\times 10^{30}$ erg (surface field). This discrepancy results from the already well-known difficulty with reconciling the magnetograms with observed open flux, which is currently not understood, leading to an inability to discriminate between these two calculated torques. The third method is based on the observed spin-rates of Sun-like stars, which decrease with age, directly probing the average angular momentum loss. This method gives $6.2\times 10^{30}$ erg for the solar torque, larger than the other methods. This may be indicative of further variability in the solar torque on timescales much longer than the magnetic cycle. We discuss the implications for applying the formula to other Sun-like stars, where only surface field measurements are available, and where the magnetic variations are ill-constrained.
\end{abstract}



\keywords{magnetohydrodynamics (MHD) - stars: low-mass - stars: stellar winds, outflows - stars: magnetic field- stars: rotation, evolution }


\section{Introduction}
Angular momentum loss through stellar winds explains the rotational evolution of low mass stars ($M_*\leq1.3M_{\odot}$) on the main sequence. These stars are shown to have outer convection zones \citep{marcy1984observations, donati2006large, morin2008stable, donati2008magnetic, petit2008toroidal, morgenthaler2011direct, gregory2012can, reiners2012observations, folsom2016evolution, folsom2017evolution}, which are able to support magnetic fields through the interplay of rotation and convection, forming a dynamo \citep{brun2017magnetism}. The magnetic field generation of such dynamos is linked with rotation \citep{browning2008simulations, reiners2009evidence, reiners2010volume, vidotto2014stellar, see2015energy, shulyak2017strong}, such that a faster rotator will, in general, produce a larger field strength. Stellar winds are found to be more effective at slowing rotation in the presence of these large scale magnetic field \citep{weber1967angular, mestel1968magnetic, keppens2000stellar, matt2012magnetic, garraffo2015dependence, reville2015effect}. Therefore, the relation of stellar rotation, magnetism and angular momentum loss leads to the convergence of rotation periods at late ages \citep{skumanich1972time, soderblom1983rotational, barnes2003rotational, barnes2010simple, delorme2011stellar, van2013fast, bouvier2014angular}.


Observations of the rotation rates of stars at different ages, and our knowledge of stellar structure, also give us direct constraints on the total external torque on the star. This value is independent from any knowledge of the physical mechanism for that angular momentum loss, but it probes only a long-time average torque (i.e., only on timescales smaller than the spin-down time, which can be in the range of tens to hundreds of Myr for main sequence stars). With the increasing number of accurate rotation period measurements available to compare with model results \citep[e.g.][]{agueros2011factory, mcquillan2013measuring, nunez2015linking, rebull2016rotation, covey2016rapidly, douglas2017poking, agueros2017setting}, we are able to examine the physical mechanisms of stellar wind braking in greater detail \citep{irwin2009ages, bouvier2014angular}.  A variety of spin evolution models have been developed to date \citep[e.g.][]{gallet2013improved, van2013fast, gallet2015improved, johnstone2015stellar, matt2015mass, amard2016rotating, sadeghi2017semi,see2018open}, which relate basic stellar properties; mass, radius, rotation period, field strength and mass loss rate, with results from analytic or numerical models for the spin down torque applied to the star, and the subsequent redistribution of internal angular momentum. 

Stellar mass and radius remain essentially constant throughout the main sequence. However, in addition to the long-time secular changes of the magnetic field due to rotation, magnetic activity is also observed to vary significantly over timescales of years to decades \citep{baliunas1995chromospheric, azizi2017survey}. This is routinely observed for the Sun which is known to have a magnetic activity cycle \citep{babcock1961topology, wilcox1972annual, willson1991sun, guedel1997x, gudel2007sun, schrijver2008global}, moving from an activity maximum through minimum and back to maximum in roughly 11 years. The Sun's cyclic behaviour is apparent in changes to the large scale magnetic field \citep{derosa2012solar}, which significantly modifies the solar wind structure and outflow properties \citep{smith1995ulysses, mccomas2000solar, wang2000long, tokumaru2010solar}. Activity cycles on other stars are quantified using activity proxies such as the long term monitoring of Ca II $HK$ emission \citep{baliunas1995chromospheric, egeland2017mount}, observed lightcurve modulation due to star spots \citep{lockwood2007patterns}, X-ray activity \citep{hempelmann1996coronal} and more recently Zeeman Doppler Imaging, ZDI \citep{semel1989zeeman, donati1989zeeman, brown1991zeeman, donati1997zeeman}. The mass loss rate of the Sun is shown to vary with the magnetic cycle \citep{mccomas2013weakest} and is fundamentally connected with magnetic activity \citep{cranmer2007self}. This behaviour is expected to be similar for other low mass stars. 




Previous theoretical studies have shown the variation in angular momentum loss over magnetic cycles \citep{pinto2011coupling, garraffo2015dependence, reville2015solar, alvarado2016simulating, reville2017global}. However they require costly MHD simulations which attempt to simultaneously fit the mass loss rate and magnetic field strengths for single epochs. In contrast, by utilsing stellar wind braking formulations from \cite{reville2015effect}, \cite{finley2017dipquad}, \cite{pantolmos2017magnetic} and \cite{finley2018dipquadoct}, hereafter FM18, which can easily predict the torque for any known mass loss rate and magnetic field strength/geometry. This allows, for the first time, a more continuous calculation of the angular momentum loss rate.

Using the multitude of current observations of the Sun (this work), and multi-epoch studies of other stars from the ZDI community (Paper II), we can now evaluate the variation of stellar wind torques over decadal timescales. We briefly reiterate the angular momentum loss prescriptions from FM18 in Section 2, collate solar observations in Section 3, and implement them in Section 4 to produce the most up-to-date determination of the solar braking torque, using methods based on the surface magnetogram data obtained from SOHO/MDI and SDO/HMI, and evaluating the open magnetic flux from the \textit{Ulysses} and the Advanced Composition Explorer (ACE) spacecrafts, along with an estimate based on the rotational behaviour of other Sun-like stars. Section 5 then discusses our result and addresses the observed discrepancy between surface field and open flux methods, along with the differences between our torque value and the derived long-time average result.

Previous theoretical studies have shown the variation in angular momentum loss over magnetic cycles \citep{pinto2011coupling, garraffo2015dependence, reville2015solar, alvarado2016simulating, reville2017global}. However they require costly MHD simulations which attempt to simultaneously fit the mass loss rate and magnetic field strengths for single epochs. By contrast, using the stellar wind braking formulations from \cite{reville2015effect}, \cite{finley2017dipquad}, \cite{pantolmos2017magnetic} and \cite{finley2018dipquadoct}, hereafter FM18, which can easily predict the torque for any known mass loss rate and magnetic field strength/geometry, without need for new simulations. This allows, for the first time, a more continuous calculation of the angular momentum loss rate.

Using the multitude of current observations of the Sun (this work), and multi-epoch studies of other stars from the ZDI community (Paper II), we can now evaluate the variation of stellar wind torques over decadal timescales. We briefly reiterate the angular momentum loss prescriptions from FM18 in Section 2, collate solar observations in Section 3, and implement them in Section 4 to produce the most up-to-date determination of the solar braking torque, using methods based on the surface magnetogram data obtained from SOHO/MDI and SDO/HMI, and evaluating the open magnetic flux from the \textit{Ulysses} and the Advanced Composition Explorer (ACE) spacecrafts, along with an estimate based on the rotational behaviour of other Sun-like stars. Section 5 then discusses our result and addresses the observed discrepancy between surface field and open flux methods, along with the differences between our torque value and the derived long-time average result.

\section{Semi-Analytic Torque Formulations}
FM18 provides semi-analytic prescriptions for the angular momentum loss rate based on over 160 stellar wind simulations using the PLUTO magnetohydrodynamics (MHD) code \citep{mignone2007pluto, mignone2009pluto}. The simulations in FM18 use a polytropic equation of state, which equates to a thermally driven wind with a coronal temperature of 1.7MK for the Sun, and a polytropic index of $\gamma=1.05$, which is nearly isothermal. The use of a nearly isothermal wind leads to some discrepancy with the observed multi-speed solar wind, which is known to be bimodal in nature \citep{ebert2009bulk}. Nevertheless, work by \cite{pantolmos2017magnetic} has shown changes to this assumed global wind acceleration can be understood within these models, and have a well described impact on our result. 

As discussed in \cite{pantolmos2017magnetic} variations in the chosen wind speed, i.e. a wind comprised of all slow or all fast wind, differ by a factor of $\sim2$ in the predicted torque. In reality the solar wind is comprised of both components, with the relative fraction of slow and fast wind changing with magnetic activity, which means the true torque is between these two extremes. For this work, we adopt the parameters derived originally in FM18, with a temperature between the extremes \citep[see][]{pantolmos2017magnetic}, and accept potential discrepancies in the wind acceleration over the solar cycle. 

The simulations of FM18 are axisymmetric, so derived torques neglect 3D effects as observed in the simulations of \cite{reville2017global}. The advantage of these formulations is that calculations can be performed much faster than MHD simulations. This allows us to use all the available data to produce the most coherent picture of solar angular momentum loss over the last 22 years. 

The torque, $\tau$, due to the solar wind is then given by,
\begin{equation}
\tau=\dot M \Omega_{*}R_*^2 \bigg(\frac{\langle R_{\text{A}} \rangle}{R_*}\bigg)^2,
\label{torque}
\end{equation}
where, $\dot M$ is the solar wind mass loss rate, the stellar rotation rate is assumed to be solid body (no differential rotation) with $\Omega_*=\Omega_{\odot}= 2.6\times 10^{-6}$rad/s, and $R_*$ is the stellar radius for which we adopt $R_*=R_{\odot}=6.96\times 10^{10}$cm. As with previous torque formulations, equation (\ref{torque}) defines the average Alfv\'en radius, $ \langle R_{\text{A}} \rangle$, to behave as a lever arm, or efficiency factor for the stellar wind in braking the stellar rotation \citep{weber1967angular, mestel1968magnetic}. 

\subsection{Formulation Using Surface Magnetic Field}
In equation (\ref{torque}) the torque depends on the average Alfv\'en radius. Simulations of FM18 showed that $ \langle R_{\text{A}} \rangle$ can be predicted using the wind magnetisation parameter, 
\begin{equation}
\Upsilon=\frac{B_*^2R_*^2}{\dot M v_{\text{esc}}},
\label{up}
\end{equation}
where the total axisymmetric field strength is evaluated using the polar field strengths from the lowest order modes $B_*=|B_*^{l=1}|+|B_*^{l=2}|+|B_*^{l=3}|$ ( $l$ is the magnetic order, for which 1, 2 and 3 correspond to the dipole, quadrupole and octupole modes respectively), and the escape velocity is given by $v_{esc}=\sqrt{2 G M_*/R_*}$, for which we adopt $M_*=M_{\odot}=1.99\times 10^{33}$g.

For mixed geometry axisymmetric fields, the average simulated Alfv\'en radius is found to behave as a broken power law of the form,
\begin{equation}
  \frac{\langle R_{\text{A}} \rangle}{R_*}=\max\Bigg\{
  \begin{array}{@{}ll@{}}
    K_{\text{dip}}[\mathcal{R}_{\text{dip}}^2\Upsilon]^{m_{\text{dip}}},  \\
    K_{\text{quad}}[(|\mathcal{R}_{\text{dip}}|+|\mathcal{R}_{\text{quad}}|)^2\Upsilon]^{m_{\text{quad}}}, \\
    K_{\text{oct}}[(|\mathcal{R}_{\text{dip}}|+|\mathcal{R}_{\text{quad}}|+|\mathcal{R}_{\text{oct}}|)^2\Upsilon]^{m_{\text{oct}}},
  \end{array}
  \label{DQO_law}
\end{equation} 
which approximates the stellar wind solutions from FM18, using $K_{\text{dip}}=1.53$, $K_{\text{quad}}=1.70$, $K_{\text{oct}}=1.80$, $m_{\text{dip}}=0.229$, $m_{\text{quad}}=0.134$, and $m_{\text{oct}}=0.087$. The variables describing the field geometry, $\mathcal{R}_{\text{dip}}$, $\mathcal{R}_{\text{quad}}$, and $\mathcal{R}_{\text{oct}}$ are defined as the ratios of the polar strengths of each mode over the total; i.e., $\mathcal{R}_{\text{dip}}=B_*^{l=1}/B_*$, etc. The scaling of equation (\ref{DQO_law}) is such that for most field strengths, in the solar case, we find that the dipole only term dominates the angular momentum loss (i.e. the dipole-only formulation of \citealp{matt2012magnetic} holds).

\subsection{Formulation Using Open Magnetic Flux}
\cite{reville2015effect} show that by parametrising the relationship for the average Alfv\'en radius in terms of the open magnetic flux, $\phi_{open}$, a scaling behaviour independent of magnetic geometry can be formulated. Such a general formula for the torque is very useful. However, the open magnetic flux cannot be observed for other stars than the Sun. We define the unsigned open flux as,
\begin{equation}
\phi_{open}=\oint_A|{\bf B}\cdot {\bf dA}|,
\end{equation}
where $A$ is a closed spherical surface located outside of the last closed field loop, i.e. in the magnetically open wind. The wind can then be parametrised with the open flux wind magnetisation, 
\begin{equation}
\Upsilon_{\text{open}}=\frac{\phi_{\text{open}}^2/R_*^2}{\dot M v_{\text{esc}}},
\label{up_open}
\end{equation}
and the average Alfv\'en radius given by,
\begin{equation}
\frac{\langle R_{\text{A}}\rangle}{R_*} = K_{\text{o}}[\Upsilon_{\text{open}}]^{m_{\text{o}}},
\label{open_law}
\end{equation}
where, from FM18,  $K_{\text{o}}=0.33$ and $m_{\text{o}}=0.371$. Here we assume the dipolar coefficients as the dipolar fraction of the total field $\mathcal{R}_{\text{dip}}$ remains significant throughout the solar cycle, with few exceptions. 

The simplicity of the semi-analytic derivation for the open flux torque formulation (see \citealp{pantolmos2017magnetic}) suggests that this method produces the most reliable torque for a given estimate of the open flux. This method is insensitive to surface geometry and any details of how the field is opened. The only factors that cause the angular momentum to deviate from this formulation is the wind acceleration and the 3D structure of the mass flux.

\section{Observed Solar Wind Parameters}
Information regarding the magnetic properties of the Sun are used here in two forms. Firstly, synoptic magnetograms of the surface magnetic field produced by both the Michelson Doppler Imager on-board the Solar and Heliospheric Observatory (SOHO/MDI) and the Helioseismic and Magnetic Imager on-board the Solar Dynamic Observatory (SDO/HMI), from which we calculate time-varying magnetic field strengths for the dipole, quadrupole and octupole field components. Secondly, measurements of the interplanetary magnetic field (IMF) strength are taken in-situ by the \textit{Ulysses} and ACE spacecrafts, which we use to produce an estimate of the time-varying solar open flux. Measurements of the solar wind speed and density are also made in-situ by multiple spacecrafts, but here we focus on results from \textit{Ulysses} and ACE. 

During the calculation of our solar wind quantities, we perform 27-day averages to remove any longitudinal variation and produce more representative values for the global wind. In doing this we have removed information of any temporal or spatial variation on smaller scales than this, which has been shown by previous authors \citep[e.g.][]{deforest2014inbound}. 

An additional complication arises from Coronal Mass Ejections (CMEs), or Interplanetary CMEs (ICMEs) as they arrive at the spacecraft detectors. ICMEs are observed in the data as impulsive increases in the in-situ solar wind properties. CMEs occur on average up to 5 times a day at solar maximum and 1 every 2-3 days at solar minimum \citep{webb2017there}. Some authors have removed these events from their datasets \citep[e.g.][]{cohen2011independency} using CME catalogues \citep[e.g.][]{cane2003interplanetary} and by identifying anomalous spikes. CMEs carry only a few percent of the total mass loss rate which is mainly located near the maximum of activity, and due to the distribution of their ejection trajectories into the Heliosphere, a reduced fraction of these events are recorded at the in-situ detectors. 

In order to gauge the impact of the enhanced magnetic field strengths and densities carried by ICMEs, we re-ran the analysis, removing periods when the wind density and field strength are greater than 10cm$^3$ and 10nT respectively from the hourly spacecraft data \citep[as done for \textit{Ulysses} by][]{cohen2011independency}. This results in $\sim3\%$ of the hourly data being cut in each 27-day average at solar maximum, and $\sim0\%$ at the minimum. During the 22 years this averages to removing $\sim1\%$ of the data from each 27-day bin. We find by removing the ICMEs the average open flux and mass loss rate we derive are reduced by $\sim4\%$. However, as CMEs should have a contribution to the total torque we prefer to include these events in our derived mass loss rate and open flux, even though there is not yet a model to show how their angular momentum loss per mass loss rate may be different than that of a steady global wind \citep[see, e.g.][]{aarnio2012mass}. As such the results presented in the remainder of this work use the full, unclipped data set. 

\subsection{Surface Magnetic Variability From SOHO/MDI and SDO/HMI}
   \begin{figure*}
   \centering
    \includegraphics[trim=0cm 0cm 0cm 0cm, width=\textwidth]{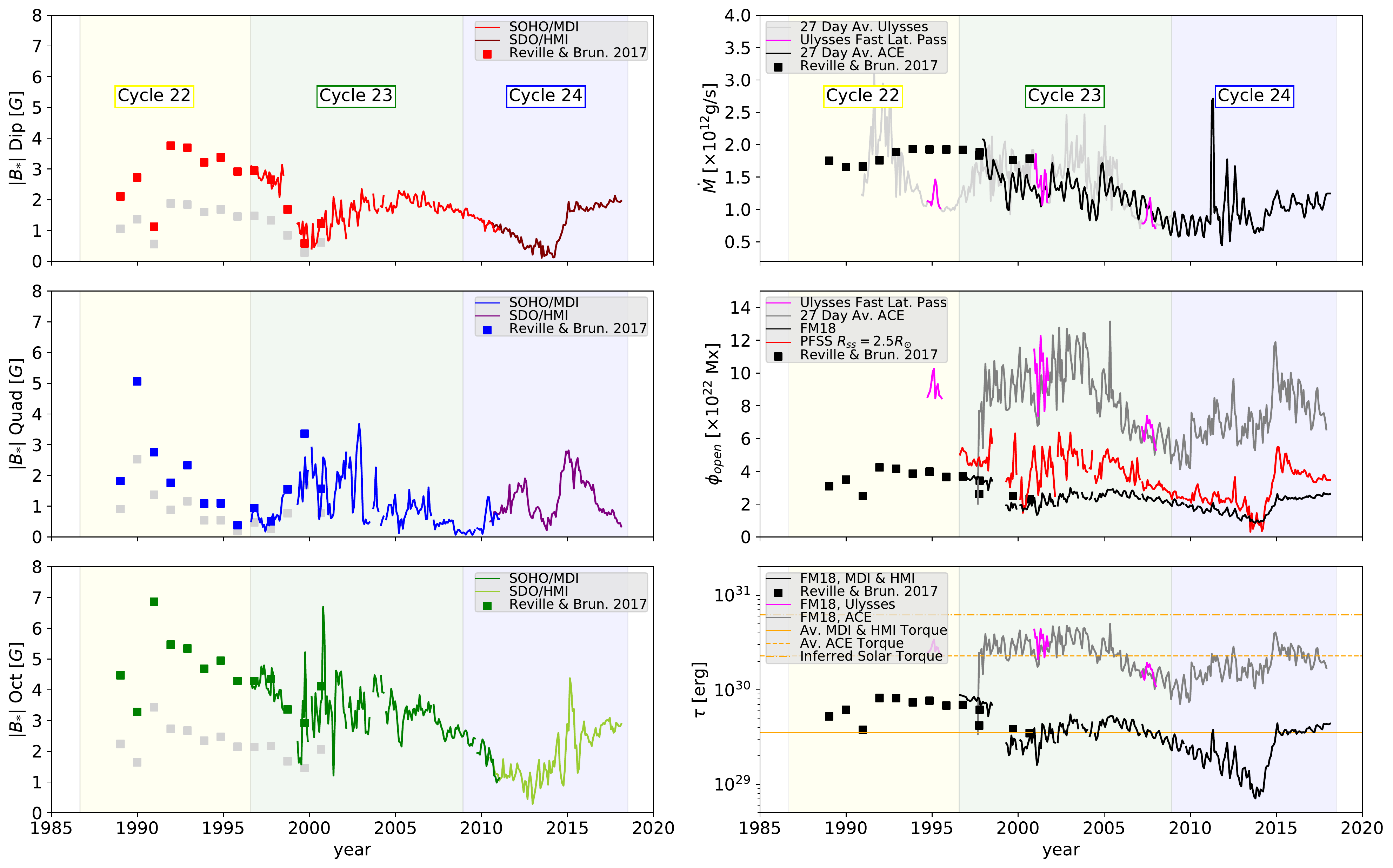}
     \caption{Calculation of the angular momentum loss rate, $\tau$, in the solar wind over the last 22 years, through various observations and utilising the torque formulations presented in FM18.  Left: Three panels presenting the lowest order spherical harmonic components, dipole, quadrupole and octupole, from the SOHO/MDI and SDO/HMI instruments along with the Carrington-rotations used in \cite{reville2017global} (grey squares are derived from WSO maps directly, where as coloured squares include a scale factor to bring the observations in line with MDI and HMI). Right: Top panel shows the mass loss rate using equation (\ref{massloss}) for, the ACE data with a black line, all \textit{Ulysses} data with a light grey line, the fast latitude scans indicated in magenta, and the models from \cite{reville2017global} with black squares. Middle panel shows the open magnetic flux using equation (\ref{openflux}) for, the ACE data with a dark grey line, \textit{Ulysses} fast latitude scans in magenta, the results of using a potential field source surface model and the associated open flux from the FM18 model on the MDI and HMI magnetograms in red and black respectively, and the models from \cite{reville2017global} with black squares. Bottom panel shows the angular momentum loss rate, with the surface field method using MDI and HMI magnetograms along with the mass loss rate from the ACE spacecraft shown with a black line, and the open flux method using ACE data and \textit{Ulysses} data in grey and magenta respectively. The average torque for each method along with the value derived in equation (\ref{eq_tausunrot}) are highlighted in solid, dashed and dot-dashed lines respectively. Results from \cite{reville2017global} are indicated with black squares in each panel. }
     \label{solar_figure}
  \end{figure*}

Using synoptic magnetograms taken from MDI and HMI\footnote{http://mdi.stanford.edu/data/synoptic.html}, the complete radial surface magnetic field strength, $B_r$, is recorded over the past 22 years for each Carrington-rotation (CR)\footnote{There are some Carrington-rotations within the SOHO/MDI sample that have missing data and as such, they are excluded from our analysis.} from July 1996 (CR 1910) to present. Both instrument teams provide polar field corrected data sets \citep{sun2011new, 2018arXiv180104265S}, accounting for projection effects on the line of sight magnetic field measurements which result in a large amount of noise at the poles, along with other affects such as the Sun's tilt angle which periodically hides these areas from view. The two instruments observed the Sun over different time periods with an overlap from the beginning of HMI in May 2010 (CR 2097) until the end of MDI in December 2010 (CR 2104). Therfore the datasets have been calibrated to produce consistent results. For this work, we apply a multiplicative factor to the HMI field strengths of 1.2, as suggested by \cite{liu2012comparison}. 

We use a total of 282 synoptic magnetograms which cover the entirety of sunspot cycle 23 (August 1996 - December 2008, CR 1913 - CR 2078), and cycle 24 up to January 2018 (CR 2199). These magnetograms are decomposed into their spherical harmonic components using the {pySHTOOLS} code \citep{wieczorek2011shtools}. The magnetograms require remapping from the sine-latitude format of the observations onto an equal sampled grid, which the code can use. Each map is then decomposed into a set of spherical harmonic modes $Y^l_m$ which have order $l=1,2,3,...,l_{max}$ (a truncation limit placed at $l_{max}=150$) and degree $-l\leq m\leq l$. This process produces complex coefficients $\alpha^l_m$, which weight each of the spherical harmonic modes,
\begin{equation}
B_{\text{r}}(\theta,\phi)=\sum_{l=0}^{l=l_{\text{max}}}\sum_{m=-l}^{m=l} \alpha^l_m Y^l_m(\theta,\phi),
\label{spherical_harm}
\end{equation}
where $\theta$ and $\phi$ represent the co-latitude and longitude of the magnetograms respectively.

This method was performed by \cite{derosa2012solar} on 36 years of observations from the Wilcox Solar Observatory and, similarly to this work, the MDI data set. The results from our decomposition agree strongly with the results presented in \cite{derosa2012solar} for the MDI observations. Appendix A contains a full breakdown of the dipole, quadrupole and octupole components we calculate.

For the calculation of the solar wind torque based on the surface field, we require the dipole ($l=1$), quadrupole ($l=2$) and octupole ($l=3$) component strengths. The pySHTOOLS code produces a strength for the axisymmetric component ($m=0$) and the subsequent non-axisymmetric components ($0<|m| \leq l$) for each harmonic order $l$. The formulation from FM18 is produced using axisymmetric simulations only, here we produce a combined field strength including all $m$, rather than neglecting the non-axisymmetric components ($m>0$). We adopt the quadrature addition of field components,
\begin{equation}
B_*^l=\sqrt{\sum_{m=-l}^l (B^l_m)^2},
\label{B_dip}
\end{equation}
where $B^l_m=\alpha^l_m \text{max}(|Y^l_m(\theta,\phi|)$, characterises the polar field strength of each mode. This results in $B_*^{l=1}$ representing a combined dipole strength using all the spherical harmonic components with $l=1$ and, $m= \{-1,0,1\}$. Similarly this is done for the quadrupole ($B_*^{l=2}$) and octupole ($B_*^{l=3}$) modes. The left 3 panels of Figure \ref{solar_figure} show how these combined dipole, quadrupole and octupole field strengths (solid lines) vary as a function of time over 22 years of magnetogram observations. 

\subsection{Mass Loss Rates and Magnetic Open Flux Variability From ACE/\textit{Ulysses}}
Along with the magnetic properties of the Sun, the mass loss rate is required to calculate the loss of angular momentum in the solar wind. The ACE spacecraft\footnote{http://srl.caltech.edu/ACE/ASC/level2/}, has been performing in-situ monitoring of the fundamental solar wind properties since its arrival at the $L_1$ Lagrangian point (on the Sun-Earth line, approximately 1.5 million km from Earth) in December 1996. A global mass loss rate is constructed from the 27-day average\footnote{An averaging period of 27-days is chosen to match the average synodic period of a Carrington rotation.} over the spacecraft data, assuming the observed solar wind flux to be characteristic of the total wind (i.e the wind is isotropically the value observed by ACE),
\begin{equation}
\dot M = 4\pi \langle R^2v_{\text{r}}(R)\rho(R)\rangle_{\text{27-day}},
\label{massloss}
\end{equation}
where $\dot M$ is the observed mass loss rate, $R$ is the radial distance from the Sun of a given observation, $v_r$ is the radial wind speed, and $\rho$ is the mass density of the wind.

The mass loss rate produced from ACE data is shown in the top right panel of Figure \ref{solar_figure} using a solid black line. The estimated mass loss rate varies between $0.43-2.72\times 10^{12}$g/s, with an average value of $1.14\times 10^{12}$g/s, which is consistent with previous works \citep{wang1998cyclic, cranmer2008winds, cranmer2017origins}. 

The same calculation is performed on the data available from the \textit{Ulysses} spacecraft\footnote{http://ufa.esac.esa.int/ufa/}, shown in light grey.  \textit{Ulysses} again made in-situ observations of the solar wind. However it took a polar orbit around the Sun with perihelion at $\approx1.35$AU (Astronomical Unit) and aphelion at $\approx5.4$AU. The spacecraft was launched in late 1990 and received a gravity assist from Jupiter in 1992 which modified the inclination of the orbit to around 80$^{\circ}$. Notably, the \textit{Ulysses} spacecraft made three fast latitude scans of the solar wind, each passing from the north pole to the south pole in approximately a year. These passes occurred between, August 1994 - July 1995, November 2000 - September 2001, and February 2007 - January 2008, which corresponds to periods of minimum, maximum and minimum solar activity respectively. These time periods are highlighted in Figure \ref{solar_figure} in magenta. 

Both sets of spacecraft observations produce agreeing $\dot M$ magnitude and variation with cycle phase although they do differ on a point-by-point basis, most notably when \textit{Ulysses} was furthest from the Sun. The ACE data is concurrent with the 22 years of magnetogram observations, and as such we use this value of the mass loss rate in future calculations.

Both spacecrafts are also capable of sampling the magnetic properties of the wind, i.e. the field direction and magnitude. Since the heliospheric magnetic field at the orbital distances of both spacecraft is thought to be predominantly open \citep{riley2007alternative, owens2011open, owens2017sunward}, these measurements allow us to make an estimate of the total unsigned solar open flux,
\begin{equation}
\phi_{\text{open}}=4\pi \langle R^2 |B_{\text{r}}(R)|_{1hr}\rangle_{\text{27-day}},
\label{openflux}
\end{equation}
where $\phi_{open}$ is the unsigned open flux and $B_r$ is the radial magnetic field strength observed by the spacecraft. The use of averaged 27-day radial field measurements, again assuming isotropy, to estimate the open flux is shown to be a good approximation, as the normalised value of the radial field, $R^2|B_{\text{r}}(R)|$, is independent of heliographic latitude \citep{smith1995ulysses}. The solar wind is found to redistribute significant magnetic flux variations due to the latitudinally directed magnetic pressure gradients formed from non-isotropy \citep{wang1995solar, lockwood2004open, pinto2017multiple}. Thus a single point measurement can be used to form a reasonable approximation of the total solar flux \citep{owens2008estimating}. 

For the magnetic field observations taken with \textit{Ulysses}, it is understood that the noise on the radial component $B_r$ will grow with distance from the Sun, such that the prediction of equation (\ref{openflux}) will become discrepant to near-earth measurements around 2AU \citep{owens2008estimating}. Therefore, we limit the open flux data used from \textit{Ulysses} to include only the fast latitude scans, at which time the spacecraft was within 2AU of the Sun. ACE, located at L1, is well within this cut-off distance, therefore a complete open flux estimate is produced for the time of its observations. The solar open flux is evaluated using equation (\ref{openflux}) and shown in the second right panel of Figure \ref{solar_figure} with a solid grey line. It is found to vary over the 22 years of observations between $2.02-13.2\times 10^{22}$Mx, with an average value of $7.98\times 10^{22}$Mx. The estimated open flux is maximum around solar activity maximum for each sunspot cycle, as with the mass loss rate.

\section{Evaluating The Solar Wind Angular Momentum Loss rate}
Here we consider three methods for determining the angular momentum loss in the solar wind. The first uses the surface magnetic field strength $B_r$ (FM18, equation \ref{DQO_law}), the second uses the open magnetic flux $\phi_{open}$ (FM18, equation \ref{open_law}), and the third calculates the expected torque on the Sun based on empirical trends in the observed rotation periods of other stars, i.e, $\tau\propto\Omega_*^3$. We aim to characterise any difference between these torque predictions, and attempt to determine the most accurate estimate of the solar wind torque and its variability.

\subsection{Torque Predictions from Surface Field Measurements}
   \begin{figure*}
   \centering
    \includegraphics[trim=3cm 0cm 3cm 0cm, width=\textwidth]{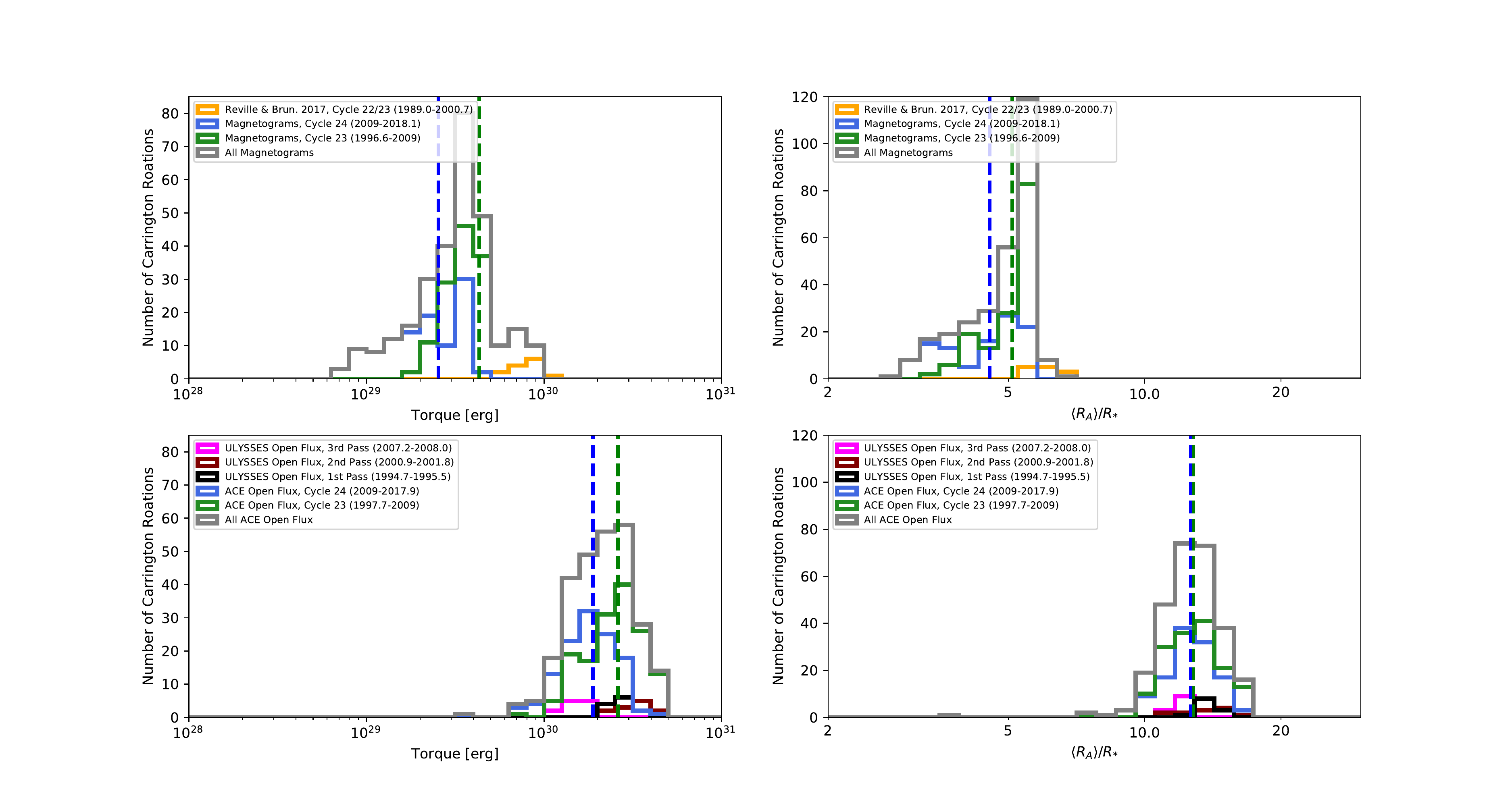}
     \caption{Histograms of the predicted torque (left) and average Alfv\'en radii (right) for the magnetogram method in Section 4.1 (top), and the open flux method in Section 4.2 (bottom). Data is binned in Carrington-rotations (27 day intervals), and coloured either green or blue for cycle 23 and 24 respectively. Cycle 23 has complete coverage, where as cycle 24 is still in its declining phase. Additionally, results from \cite{reville2017global} (yellow) are compared with the magnetogram results in the top panels. The distributions appear approximately log-normal for each cycle. In both methods the torque distribution produced is lower in the current cycle (irrespective of the incomplete data) than cycle 23 (and note the averages for the current cycle are expected to decrease as the cycle moves into an activity minimum), indicating significant variability between cycles.}
     \label{solar_histogram}
  \end{figure*}

Using the decomposed surface magnetograms from SOHO and SDO, along with mass loss rate measurements from the ACE spacecraft, we evaluate the solar wind torque using work from FM18. As previously discussed, equation (\ref{DQO_law}) applies for axisymmetric combinations of dipole, quadrupole and octupole fields only. Despite this, we assume this relations holds for the non-axisymmetric field components also, and describe a single field strength for each harmonic mode $l$ using equation (\ref{B_dip}), this assumption is discussed in Section 5.1. 


The predicted torques based on magnetograms from MDI and HMI are displayed in the bottom right panel of Figure \ref{solar_figure} with a solid black line. The average torque predicted over the 22 years of data using this method is $3.51\times 10^{29}$erg. Splitting this time period into separate sunspot cycles\footnote{The first reversed-polarity sunspot of cycle 24 occurred in January 2008, but we adopt the time of minimum smoothed monthly sunspot number which occurred in December 2008.} we produce a histogram of the torque and average Alfv\'en radii, top panels of Figure \ref{solar_histogram}. The average angular momentum loss for cycle 23 is $4.27\times 10^{29}$erg, where as cycle 24 currently has an average of $2.51\times 10^{29}$erg, $41\%$ lower. The Alfv\'en radii predicted by equation (\ref{DQO_law}) are also shown to be distinctly different for each cycle.

\subsection{Torque Predictions from Open Flux Measurements}
The angular momentum loss rates calculated using equation (\ref{open_law}) with the open flux and mass loss rate observations from \textit{Ulysses} and ACE are displayed in the bottom right panel of Figure \ref{solar_figure} in magenta and grey respectively. Notably, both sets of observations agree well, and indicate the solar maximum coincides with the maximum braking torque in the solar wind. The average torque predicted using the open flux method is $2.28\times 10^{30}$erg, which is 3.26 times greater than the surface field method in the previous section. The sunspot cycles can again be distinguished, the torque and Alfv\'en radii predicted by equation (\ref{open_law}) are shown in the bottom panels of Figure \ref{solar_histogram}. The average angular momentum loss for cycle 23 is $2.60\times 10^{30}$erg, where as cycle 24 currently has an average of $1.88\times 10^{30}$erg, $28\%$ lower. Similarly to the torque predicted from the surface field measurements, we find a decreasing value of the solar torque however the difference in average Alfv\'en radii from cycle to cycle is far smaller for the open flux method. 

\begin{figure*}
   \centering
    \includegraphics[width=0.9\textwidth]{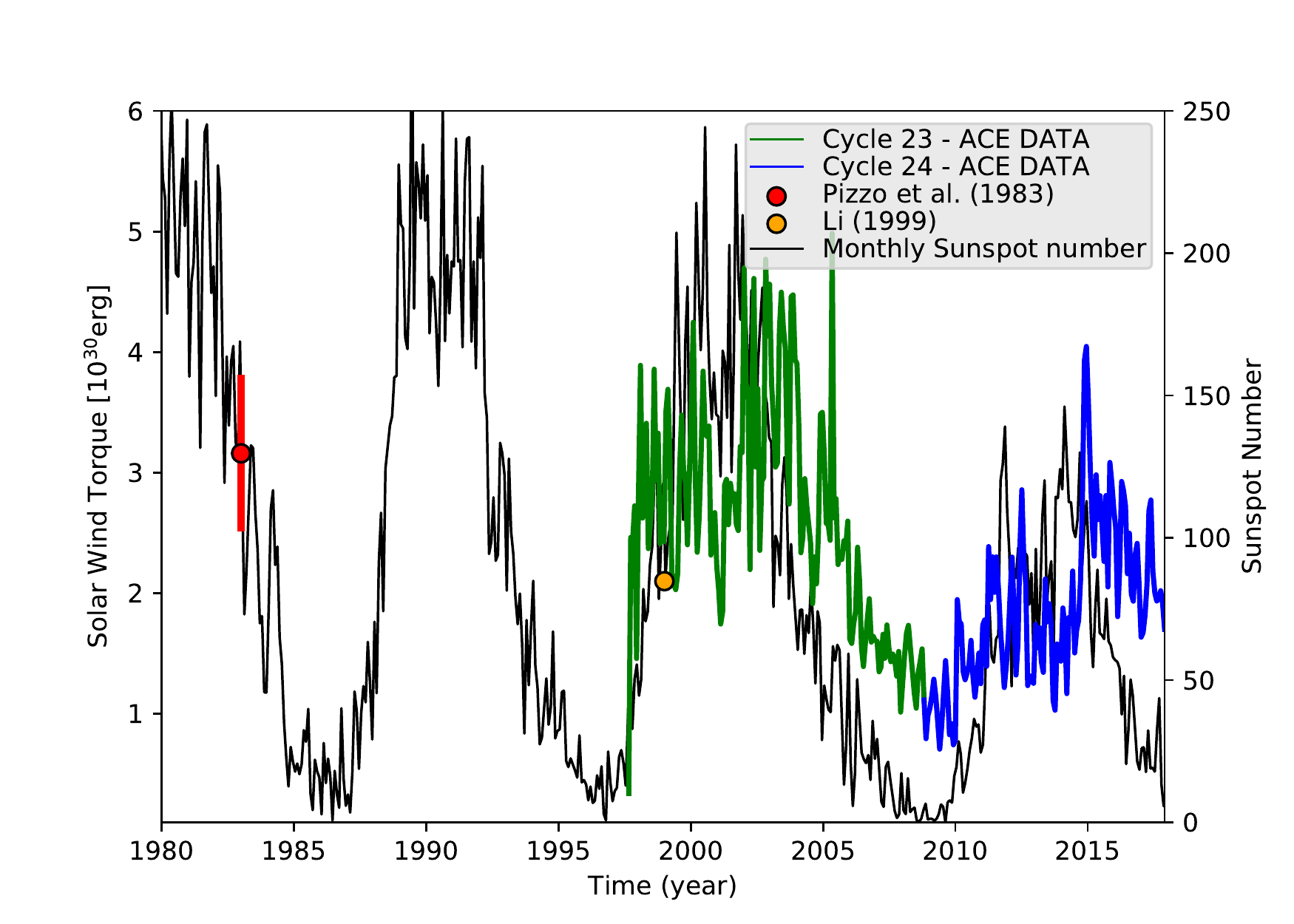}
     \caption{Solar angular momentum loss calculated using data from the ACE spacecraft and the open flux torque formulation from FM18 vs time (coloured line), along with the sunspot number (black line) and previous estimates of the angular momentum loss rate. Data from Cycle 23 is coloured green and the current Cycle 24, blue. Cycle 24 is weaker in both activity and the predicted torque. This could explain the larger value from \cite{pizzo1983determination}, which was measured during a stronger magnetic cycle.}
     \label{sunspot_torque}
  \end{figure*}

Figure \ref{sunspot_torque} displays the ACE derived torque along with the monthly averaged sunspot number; roughly the angular momentum loss rate rises in accordance with the sunspot number with a hint that it lags behind in the declining phase of the sunspot cycle (see Section 5.4 for further discussion).  

\subsection{Solar Torque Inferred from Observed Stellar Rotation Rates}
At the start of the main sequence phase, stars with nearly a solar mass, exhibit a wide distribution of rotation rates, which is observed to converge toward a narrow distribution of rotation rates by an age of a few hundred Myr \citep[e.g.][]{bouvier2014angular}. The distribution of spin rates continues to narrow, as the average spin rate decreases in time.  The narrowing distribution and common evolution of spin rates suggests that the stellar wind torques for all (or most) solar-mass stars approach a single relationship that depends simply on stellar parameters and spin rate, and which becomes independent of the ``initial'' conditions (e.g., independent of whether the star was a fast or slow rotator at the start of the main sequence phase).

This late-time, asymptotic behavior of observed stellar spin rates gives us constraints on the external torques, completely independently from any knowledge of the physics of stellar winds.  To derive such a constraint for solar-mass stars older than a few hundred Myr (following \citealp{schatzman1962theory}; \citealp{durney1985theories}; \citealp{kawaler1988angular}; \citealp{matt2015mass}), we first approximate that the external torque depends simply on the rotation rate as a power-law,
\begin{eqnarray}
\label{eq_taurot}
\tau_{\rm rot} = \langle\tau_\odot\rangle_{\rm rot} \left({\Omega \over \Omega_\odot}\right)^{p+1},
\end{eqnarray}
where $\langle\tau_\odot\rangle_{\rm rot}$ is the current long-time-average torque of the Sun, $\Omega$ is the stellar rotation rate, $\Omega_\odot$ is the solar rotation rate, and $p$ will be constrained by observations and theory.  Next, if we assume that the moment of inertia of stars is constant during the main sequence phase and that the stars rotate as solid bodies, we can integrate the angular momentum equation analytically.  This analysis shows that, for any value of
$p>0$, and independent of any reasonable ``initial'' spin rate (at young ages), the rotation rates will converge toward the relationship
\begin{eqnarray}
\label{eq_omegaconverged}
{\Omega \over \Omega_\odot} = \left({I_\odot \Omega_\odot \over p \langle\tau_\odot\rangle_{\rm rot}}{1 \over t}\right)^{1 \over p}
\end{eqnarray}
where $I_{\odot}$ is the solar moment of inertia and $t$ is the current age of the star.  From the age of a few hundred Myr to that of the Sun, it has been long known that the average spin rates of Sun-like stars decreases as approximately the inverse square root of the age \citep{skumanich1972time, soderblom1983rotational, barnes2003rotational}, which implies $p \approx 2$.

Using solar parameters, equation (\ref{eq_omegaconverged}) predicts the present-day average torque, which is required to explain observed spin rates of Sun-like stars,
\begin{eqnarray}
\label{eq_tausunrot}
\langle\tau_\odot\rangle_{\rm rot} = 6.2 \times 10^{30} \; {\rm erg} 
      \left({I_\odot \over 6.90 \times 10^{53}\; {\rm g \; cm^2}}\right)  \nonumber \\
      \times \left({\Omega_\odot \over 2.6 \times 10^{-6}\; {\rm rad/s}}\right) 
      \left({4.55 \; {\rm Gyr} \over t_\odot}\right) 
      \left({2 \over p}\right),
\end{eqnarray}
where we have input fiducial values for the solar moment of inertia \citep{baraffe2015new}, representative rotation rate \citep{snodgrass1990rotation}, age \citep{guenther1989age}, and $p$. Appendix B discusses the validity and uncertainties associated with the prediction of equation (\ref{eq_tausunrot}).

Using observed stellar rotation rates to probe the torque is only sensitive to torques averaged over some timescale that is much shorter than the spin-down time, but larger than magnetic cycle timescales.  For the ages near the Sun's, this means that equation (\ref{eq_tausunrot}) estimates the torque as averaged over a timescale of $\sim 100$ Myr.  Although the converging of stellar spin rates at late times suggest that the torques are ``well-behaved'' and predictable, the observations do not rule out that stellar wind torques could (and apparently do) vary quite substantially on short timescales.

\subsection{Comparison to Previous Calculations of the Solar Torque}
  \begin{table}
\caption{Solar Angular Momentum Loss Rates From This Work and Others}
\label{solarEstimates}
\center
\setlength{\tabcolsep}{1pt}
    \begin{tabular}{c|c}
        \hline\hline
$\langle \tau\rangle$ [$\times 10^{30}$erg]	 	&	Citation	\\	\hline
$0.35$	 &	This Work, Magnetograms / equation (\ref{DQO_law})	\\
$2.28$	&	This Work, Open Flux / equation (\ref{open_law})	\\ 
$6.20$	&	This Work, Observed Spins/ equation (\ref{eq_tausunrot}) \\ \hline
$2.51-3.77$	&	\cite{pizzo1983determination}	\\
$2.1$	&	\cite{li1999magnetic}	\\
$2.18$	&	\cite{pinto2011coupling}	\\
$0.9-2.3$	&	\cite{pantolmos2017magnetic}	\\
$0.80$	 &	\cite{reville2017global}	\\ \hline

    \end{tabular}
\end{table}

A large number of solar wind models exist in the literature, many of which produce estimates for the solar mass loss and angular momentum loss rates \citep{usmanov2000global, riley2001empirically, pinto2011coupling, alvarado2016simulating, garraffo2016missing, reville2017global, pognan2018solar}. The reported values have a wide range due to the large differences in input physics, such as the use of polytropic winds or the inclusion of a specific coronal heating function. As the mass loss rate is typically evolved self consistently in these models, differences in the modelled torque value is often due to discrepant mass loss rates when compared to observations (as this is a challenging problem). With the correct adjustments to recover the observed solar mass loss rate, most models produce a comparable value to the present work. Unlike the works above, the parametrisation of FM18 allows for the impact mass loss and the magnetic field to be decoupled such that we can produce a semi-analytic result which matches the observed solar values. 

In this section we focus on a few theoretical models, which concider the effects of magnetic variability over the solar cycle, plus data driven models of the solar wind. This includes the dynamo driven wind simulations of \cite{pinto2011coupling}, the recent 3D wind simulations of \cite{reville2017global}, and estimates for the torque using observed values such as \cite{pizzo1983determination} using the HELIOS spacecrafts and \cite{li1999magnetic} who further supported this value with data from \textit{Ulysses}. From these authors, only \cite{pinto2011coupling} and \cite{reville2017global} consider the variability of the Sun. Table \ref{solarEstimates} collects previous estimates of the solar torque and compares them to this work. 

Estimates made of the solar wind torque from \cite{pizzo1983determination} and \cite{li1999magnetic} both agree in magnitude with the open flux estimate performed in Section 4.2. \cite{pizzo1983determination} made a direct measurement of the solar angular momentum flow, which should be the most accurate method, however they required very significant assumed spacecraft pointing corrections. Therefore it is not clear how rebust the measurement is. Based on our observed variability with sunspot cycle it is expected that the estimate made for the torque using the HELIOS spacecrafts should be higher than our current average, as the Sun was more active during cycle 21, which is calculated to be in the range of $2.55-3.77\times 10^{30}$erg by \cite{pizzo1983determination} (see Figure \ref{sunspot_torque}). The average value of the solar wind torque during cycle 23 and 24 is $\sim30\%$ lower than this, which is potentially evidence for continued variability on longer timescales than considered within this work.

\cite{pinto2011coupling} used a solar-like kinematic dynamo model to drive an axisymmetric MHD wind simulation. The results did not intend to model the actual Sun, but this was the first work to include the effect of magnetic variability in the calculation of the angular momentum loss rate. Results from this work agree with the average Alfv\'en radii predicted by equation (\ref{DQO_law}) from FM18, apart from at their minimum of activity in which the axisymmetric dipole decreases without a rise in the equatorial dipole component to maintain the size of the Alfv\'en radius. Due to this, the torque predicted is strongly anti-correlated with the solar activity cycle. This highlights the need to include the equatorial component to produce a smoothly varying torque, as seen with in the open flux method. Their average torque agrees with previous works but requires a mass loss rate which is twice as large as the observed solar value. 

\cite{reville2017global} compute 13 3D MHD simulations of the solar wind, stretching between cycle 22 to cycle 23, which we compare to our results in Figure \ref{solar_figure} (shown in filled squares). Surface magnetic field data is gathered from the Solar Wilcox Observatory (WSO, \citealp{scherrer1977mean}) synoptic maps (R\'eville, private communication), which display similar trends to the MDI and HMI data used in this work, see the top three panels (grey squares). The WSO observations are known to be less sensitive and under-represent the strength of the field when compared with MDI and HMI results \citep{derosa2012solar, riley2014multi}, a multiplicative factor has been used to scale the strengths (coloured squares); note that this was not done for the values used within \cite{reville2017global}, only for completeness here. The PLUTO code is used to construct 3D wind solutions for each WSO magnetogram, this produces global values for the mass loss rate and open magnetic flux (R\'eville, private communication), which are used to generate a torque in the bottom panel of Figure \ref{solar_figure}. The mass loss rates match the observed average, with slight variation due to differences in the expansion of the field lines for the changing magnetic topolgies (also discussed in \citealp{reville2016age} and \citealp{finley2017dipquad}). However the factor of $\sim2$ variation in mass loss rate with magnetic cycle phase is not reproduced, shown in both the observations from ACE and \textit{Ulysses}. As discussed by \cite{reville2017global}, this occurs due to the lack of additional wind driving physics which should correlate with surface magnetic field energy, such as energy deposition through Alfv\'en wave heating \citep[e.g.][]{cranmer2007self, pinto2016flux}. 

Interestingly, these 3D simulations contain the non-axisymmetric components of the magnetic field ($l_{\text{max}}=15$). Since the resulting torques from these simulations and our use of equation (\ref{DQO_law}) appear to agree, it strengthens our previous assumption for including the non-axisymmetric components in our calculation. The results for the Alfv\'en radii also show good agreement, as shown in FM18. The methods used in the MHD wind simulation of \cite{reville2017global} are similar to that of FM18, using a polytropic wind acceleration profile. Therefore the torque estimate is similar to the magnetogram-based calculation from Section 4.1. 

\section{Discussion}
Using the torque formulations from FM18, the value of the solar wind torque is shown to be lower than the empirical estimate based on the rotation of other Sun-like stars. We also find a disagreement between the two predictions from FM18, using either the surface or open flux method for calculating the torque. Both methods show the angular momentum loss rate to be variable in time, seemingly linked with the strength of magnetic activity on the Sun. Differences in the dynamical torque estimates for the current Sun and the long-time-average value may then be due to magnetic variation on longer timescales than the 22 year magnetic cycle. Here we discuss such factors which may be responsible for discrepancies in our predicted torques, and also their implications for using the work of FM18 on other Sun-like stars. For which, we can only obtain basic information about their surface magnetic field. Knowing that the open flux method is perhaps the most reliable, how can we reconcile our results for future use of the surface field method?

\subsection{The Impact of Non-axisymmetric Magnetic Components}
In our calculation of the solar torque, based on surface magnetogram observations, we include the strength of the non-axisymmetric components through equation (\ref{B_dip}) which adds the components in quadrature to produce a combined strength for each mode $l$. This is done because the non-axisymmetric components of the field will impact the radial decay of the magnetic flux in a similar way to the axisymmetric components, which is the most significant driver of the location of the Alfv\'en radius (see discussion within FM18). MHD modelling by \cite{garraffo2016missing} shows the torque generated by pure non-axisymmetric geometries are comparable with their axisymmetric counterparts, which supports our assumption here. However they do not disentangle the effect of mass loss rate and magnetic field geometry/strength on their angular momentum loss rates. It therefore remains to be shown if, or how, non-axisymmetric modes change the fit parameters $K$ and $m$ in equation (\ref{DQO_law}) from FM18.

The torque calculated in Section 4.1 is controlled largely by the combined dipole field strength, which appears to be out of phase with solar activity, displayed in the top left panel of Figure \ref{solar_figure} (note the use of absolute magnitude field strengths). During each sunspot cycle, the torque is maximised at the start of the sunspot cycle and just after the dipole polarity reversal. These are in general, times when the axisymmetric dipole is maximised (see Appendix A). This behaviour is markedly different to the open flux method which clearly shows the largest angular momentum loss rate at the sunspot maxima, when the equatorial dipole component is strongest. 

In order to assess the impact of including the non-axisymmetric components with equation (\ref{B_dip}), we performed the torque analysis using both, only the axisymmetric components, and the combined strength approach of equation (\ref{B_dip}). We find marginally differing results for both approaches, most notably, using only the axisymmetric component leads to a deeper minimum field strength during the dipole polarity reversal than in the combined approach and a far lower value for the torque during this time. This is the picture presented in \cite{reville2015effect}, in these simulations the equatorial dipole component is ignored, however this component does not vanish at maximum and should impact the angular momentum loss rate. 

It is certain that the non-axisymmetric field components will contribute to the open flux in some way, and perhaps it is their relation to the torque which will resolve the discrepancy in how the torque varies over the cycle between the surface field and open flux methods. To first order, we believe our method produces more realistic results than simply taking the axisymmetric components alone. But the impact on wind acceleration and the effectiveness of the magnetic braking from these components is not completely understood.

\subsection{The Impact of Model Wind Parameters}
The FM18 model uses a particular set of fit parameters in equation (\ref{DQO_law}) which are taken from simulations using a single polytropic wind temperature. Here we assume this to be an average of the slow and fast solar wind flows. However, work by \cite{pantolmos2017magnetic} indicates this assumption produces a form of uncertainty since we do not know the best fitting temperature for the Sun (and especially not for other stars). It is likely that the correct average polytropic wind temperature is also slightly variable during the solar cycle, with a differing ratio of fast and slow wind present. In general, variability in the wind temperature over the cycle will affect both torque formulas from FM18 and so represents an uncertainty on our results, i.e. for a fixed $\dot M$, a faster wind will open more flux with a weaker resulting torque.

Differences in the observed variability of the solar wind torque between FM18 methods, i.e. the open flux method being smoothly varying and the surface field method being heavily dependent on the input dipole field strength, may be explained by the 3D and multi-speed nature of the solar wind. Here we have assumed for the surface field method, that the non-axisymmetric components will contribute to the torque though a quadrature addition of their strengths with the axisymmetric field, see equation (\ref{B_dip}). However, their relationship may be more subtle and interconnected with the wind acceleration, in effect smoothing the variability of the torque over the cycle. Models of the solar wind which recover the bi-modality of wind properties \citep{alvarado2016simulating}, such as those produced for space weather prediction \citep[e.g.][]{usmanov2000global, toth2005space}, as of yet have not been used to formulate a useable scaling relation for how the solar wind angular momentum loss rate scales with various parameters, such as mass loss rate or magnetic field strength.

\subsection{The Open Flux Problem}
Synoptic magnetograms are produced from a wide range of observatories, both in space and on the ground, for which line of sight magnetic field measurements are processed using different methods into coherent pictures of the whole solar surface. These magnetic maps agree qualitatively, with the same morphology of active regions and distributed surface flux. However, they often disagree quantitatively requiring saturation/correction factors to be brought into agreement with one another \citep{wang1995solar, liu2012comparison, riley2014multi}.

Commonly Potential Field Source Surface (PFSS) models are used with these magnetogram observations as an input boundary condition, which allows for a quick and qualitative view of the coronal magnetic field. However, these PFSS models either under-estimate the solar open flux with a source surface around $2.5R_*$, but match the observed coronal hole morphology and area, or require much smaller source surface radii $<2R_*$ to match the observed solar wind open flux at the cost of over predicting the coronal hole area \citep{riley2006comparison, lee2011coronal, arden2014breathing,reville2015solar, reville2017global, linker2017open}. 

The surface flux and open flux methods from FM18 disagree in their prediction of the average torque over the last 22 years. The value derived based on the observed open flux in the solar wind is $\sim7$ times larger than the value produced using the magnetogram observations. The main disagreement between these two approaches centres on the amount of open flux produced from the magnetograms. Equating equations (\ref{DQO_law}) and (\ref{open_law}), and solving for $\phi_{open}$, we can produce a relation for the open flux predicted by the FM18 models, given a surface field strength,
\begin{equation}
\phi_{open} =\bigg[ \bigg(\frac{\langle R_A\rangle}{R_*}\bigg|_{eq. 3}\frac{1}{K_o}\bigg)^{1/m_o} R_*^2\dot{M} v_{esc}\bigg]^{1/2}
\label{invert}
\end{equation}
where ${\langle R_A\rangle}/{R_*}|_{eq. 3}$ is our predicted Alfv\'en radii given by equation (\ref{DQO_law}). This corresponds to an average open flux of $2.21\times 10^{22}$Mx from the magnetograms, a factor of 3.61 lower than is observed by ACE. This is shown through the full dataset with a solid black line in open flux panel of Figure \ref{solar_figure}. We also produce an estimate of the open flux using a PFSS model \citep{altschuler1969magnetic} with a constant source surface radius of 2.5$R_{\odot}$, shown with a solid red line. The PFSS produces a similar magnitude of open flux to FM18 with some differences, both systematically under-predicting the observed open flux. Differences between these models are undoubtedly linked to the FM18 models predicting the coronal magnetic field becoming radial/open at much larger distances, than the fixed PFSS source surface of $2.5R_{\odot}$, during that time. 

The discrepency when extrapolating the solar open flux from magnetogram observations has been a persistent issue in the solar community \citep{zhao1994coronal, wang2000long, lockwood2004open, stevens2012underestimates}. For example, the Space Weather Modelling framework \citep{toth2005space} require the input magnetograms to be scaled by a factor of 2-4 to improve the comparison with observations \citep{cohen2006semiempirical, oran2013global, pognan2018solar}. It has been suggested that the addition of shear and twisting of magnetic field lines can allow more open flux, but again this impact the coronal hole area predicted from the models \citep{edwards2015influence}. It is therefore generally accepted that magnetograms require multiplication by an uncertain factor or the inclusion of additional magnetic flux (typically coronal mass ejections or small scale surface fields) in order to bring observations in-line with the extrapolated field strength at 1AU \citep{wang1993latitude, zhao1995prediction, cohen2006semiempirical, riley2007alternative, riley2014multi}.


Authors such as \cite{lowder2017coronal} using the OMNI database and \cite{owens2008estimating} using historical heliospheric spacecraft, obtain values in agreement with our ACE 27-day averages. However, another source of disagreement in the open flux may come from these observed values. \cite{lockwood2009excess} suggest that using the IMF measurements to infer the solar open flux may lead to overestimation. This is due to longitudinal structures in the solar wind where the IMF twists back on itself and therefore increases the observed flux passing over the spacecraft \citep{crooker2004large}. The actual impact of this effect and others on our measured open flux value is uncertain.  In order to take advantage of the open flux torque formulation from FM18 and other previous works, accurate observations of the solar open flux are required, which will likely occur with the launch of both Parker Probe \citep{fox2016solar} and Solar Orbiter missions \citep{mueller2013solar}. Previous estimates of the solar Alfv\'en surface height, based on observed solar wind properties, place the minimum average Alfv\'en radius around $10-15R_*$ \citep{zhao2010magnetic, deforest2014inbound}. These values appear most consistent with our calculation using the open flux, presented in Section 4.2. Because there also appears to be fewer uncertainties, we assume the open flux torque is the most reliable result, however more work is needed.

\subsection{Torque Variability Over Magnetic Cycles}
Section 4 discussed the average value of the solar torque. However, our interest in using this dataset is primarily motivated by the variability of the angular momentum loss rate with solar cycle phase. Here we focus on this variation and the differences in the torque between sunspot cycles 23 and 24. 

The angular momentum loss rate predicted from the magnetograms with equations (\ref{torque}) and (\ref{DQO_law}) is heavily dependent on the dipolar component of the global magnetic field. The same is not observed for the open flux formulation, which smoothly varies over the cycle but is in general positively correlated with dipole component. The open flux torque becomes largest around periods of sunspot maximum, at which point the equatorial dipole field strength and open flux are known to be maximised \citep{wang2002sunspot}, hence why previously we included the non-axisymmetric components in our torque calculation. Despite our inclusion of the non-axisymmetric field strengths with equation (\ref{B_dip}), the surface field method does not produce a smoothly varying angular momentum loss rate with solar cycle. Further work is required to resolve this, potentially by employing better thermodynamics in the wind and a treatment for the non-axisymmetric components.

The angular momentum loss rates and average Alfv\'en radii determined from both methods are binned into 27-day (carrington rotation) averages, and presented in histograms in Figure \ref{solar_histogram}. The discrepant average torque values are evident between the surface field and open flux methods. Further, data is coloured by sunspot cycle, as done in Figure \ref{sunspot_torque}, with cycle 23 and 24 in green and blue respectively. Our dataset spans the entirety of cycle 23 and the majority of cycle 24. Each cycle is observed to broadly follow a log-normal distribution. Both methods concur that cycle 24 has a lower average torque and Alfv\'en radius than cycle 23, (see vertical dashed lines for averages). The averages of the Alfv\'en radii predicted from the open flux method are nearly constant between cycles, but as cycle 24 is currently moving into a minimum the average is expected to move lower as it becomes complete. Viewing the \textit{Ulysses} fast passes, the average Alfv\'en radii for the minimum of cycle 24 (3rd pass) is smaller than that of the minimum of cycle 23 (1st pass), supporting this hypothesis. 

Despite the discrepancy in magnitude between both methods for determining the torque, the variation between cycles shows a similar trend. This implies the surface flux formulation can be brought into rough agreement with the open flux technique using a multiplicative scaling factor, which has been done previously to match observed spin-evolution distributions \citep{gallet2015improved, amard2016rotating, sadeghi2017semi, see2018open}. 

\subsection{Long-term Torque Variability}

The torque from both the surface field and open flux methods are shown to be variable with magnetic cycle, and appear to be decreasing from cycle 23 to 24. This work investigates the angular momentum loss rate over decadal timescales, but the process of rotational evolution is known to occur over billions of years. In which case, it is hard to tell if the current solar wind torque is typical of the long-time average value. 

The torque in the solar cycle here varies by a factor of $5-10$ and the average is a factor of $2.7-18$ below the inferred torque from Section 4.3, depending on the method used (open-surface). Perhaps one way to reconcile them is if the solar wind torque varies with a much larger amplitude on a timescale much longer than 20 years, as probed here, but less than the timescales probed by observations of stellar spin rates of $\sim$100 Myr.  I.e.,   It is possible that the solar wind torque is currently in some kind of ``low state'' relative to the long-time average. For example, if our present-day torque is a factor of about 4 smaller than the long-time average, to recover the average this implies that the Sun should either spend substantially longer in a slightly higher activity/torque state than the current low state, or spend an equivalent amount of time as the current low state with a torque 7 times bigger than present. In this extreme case, it requires a dipole field strength about 8 times bigger or mass loss rate 30 times bigger (or some combination of the two). If this is true, we should see Sun-like stars with the same rotation rate as the Sun, but with on average more magnetic activity than the Sun (the Sun should be below average for its Rossby number).

Using activity proxies magnetic variability is recovered on timescales of centuries \citep{lockwood2007patterns, lockwood2009rise}. Models of the solar open flux from \cite{vieira2010evolution} show that the Sun is at a low in open flux, but the current value is not exceptional. With the open flux scaling almost linearly with the torque predicted from FM18, averaging the torque on longer $\sim100$yr timescales could increase the predicted value towards agreement with the inferred torque from Section 4.3. 

Additionally, \cite{van2016weakened} suggests a transition around the solar age to a weakened form of magnetic braking. They suggest the braking torque becomes weak enough to be insignificant for subsequent main-sequence evolution of rotation, requiring a very sharp reduction in the braking torque. The smaller value of torque found in other sections could be interpreted as agreement with the \cite{van2016weakened} hypothesis, although it seems unlikely we live in a time immediately following such a transition.  The discrepancies presented here from our predicted long-time average can seemingly be explained by many other factors, such as our chosen wind temperature or long-time variability, so this appears coincidental.

\subsection{Application to other Sun-like Stars}

Our position on Earth is unique for observing the solar wind. We are embedded in the expanding solar atmosphere, and as such we can access both in-situ observations of the basic solar wind properties, and take advantage of remote sensing to build an accurate picture of solar magnetism using a variety of telescopes. This work utilises this wealth of data available for our local star, which we have an almost complete coverage with a monthly cadence for 22 years. 

For other stars this is not possible as the tenuous emission of their stellar winds is undetectable. In order to gain information about the mass loss rate and wind properties of these distant stars, we rely on proxies such as the strength of Lyman-$\alpha$ absorption at their astropauses \citep{wood2004astrospheres} and more recently the observed erosion of exoplanet atmospheres \citep{vidotto2011prospects, vidotto2017exoplanets}. The magnetic field topology and strength of Sun-like stars are sampled using techniques such as Zeeman broadening and Zeeman Doppler Imaging, which at best, produce a measurement of the stellar magnetic field on yearly timescales \citep{morgenthaler2012long, jeffers2014e, saikia2016solar}. This leads to the question, how does our ability to measure the mass loss rate and magnetic field impact the predictions of their stellar wind torques?

While performing this analysis on the Sun, we have gained some insight into these effects: 
\begin{enumerate}
\item Torques derived using stellar magnetic field observations and equation (\ref{DQO_law}) may be lower than in actuality, due to the FM18 model producing a smaller value of unsigned open flux than measured in the solar wind. It remains to be shown if this can be corrected for by the application of a common scaling factor (this work indicates $\sim 15$). 
\item Magnetic variability can lead to estimates of the angular momentum loss which are, in the solar case, up to a factor of $\sim 10$ different from one observation to another. Observations of other Sun-like stars will therefore suffer from considerable uncertainty in their derived angular momentum loss rates based on a single or small number of observations. 
\item Long-time variability may also play a role, and with the difficultly ascertaining the true magnetic behaviour of other Sun-stars, i.e. if they are cyclic or stochastic, the corresponding estimate of their angular momentum loss rate may be discrepant from rotation evolution model predictions.
\end{enumerate}

Paper II, and also See et al. (in Prep), focus on applying the formulations of FM18 onto Sun-like stars for which we have information on their magnetic topology and variability. Again these results are compared to predictions from spin-evolution modelling \citep{matt2015mass}, using the information gained here from the Sun to interpret the results.

\section{Conclusion}

In this work we have utilised the wealth of current solar observations and the semi-analytic results from FM18 to produce an estimate of the current solar wind torque. This is compared with spin evolution calculations and shown to be a factor of 2-3 smaller than expected.

Two angular momentum loss prescriptions from FM18 are implemented using observed surface field strengths from the SOHO and SDO spacecrafts, along with mass loss rates and open flux measurements from \textit{Ulysses} and ACE spacecrafts. The methods are found to produce average torques which either differ due to the amount of unsigned open flux the FM18 model produces from a given surface field observation, or the potential over-prediction of the open flux from spacecraft measurements. Assuming the open flux measurements from the in-situ spacecrafts are valid, we predict the solar wind torque has a present-day value of $2.3\times 10^{30}$erg, averaged over the last 22 years.

The observation that the spin rates of Sun-like stars converge toward a single track that depends on age also allows us to derive equation (\ref{eq_omegaconverged}), describing how this spin-down depends on the torque and stellar properties. Then using solar parameters in equation (\ref{eq_tausunrot}) predicts that the long-time averaged torque should be $6.2\times 10^{30}$erg. Comparing this estimate of the torque from observed spin-evolution to the present-day torques predicted by the dynamical models gives additional insights.  Differences in the average present-day torques to the spin-evolution torques, could be due to, (a) variability on a longer timescale than probed by the present-day variability presented here (but less than a spin-down time), (b) errors in using the dynamical models inferring present-day torque, or (c) that stars spin-down significantly different than Skumanich at ages of a few to several Gyr.  We need additional information to discriminate between these possibilities. The required variability of (a) suggests we should observe stars like the Sun that are on average significantly more active (i.e., that they have larger torques) such that the average is correct. From the dynamical models, (b), uncertainties remain in the wind acceleration and effects of non-axisymmetric field components which both require further study to disentangle. Observationally, (c) requires more period-mass-ages for old stars to confirm or refute the \cite{van2016weakened} hypothesis.  

For other Sun-like stars, measurements of their unsigned open flux and mass loss rates are not readily available. Instead we rely on surface magnetic field measurements, which are gained through Zeeman Doppler Imaging and Doppler Broadening techniques. Using the FM18 formula, predictions of the angular momentum loss rates for these stars based on their surface measurements may be smaller than in reality. Future models should be refined to better match the wealth of solar data available, such models should be able to open the correct amount of flux from a given surface magnetic field observation and continue to remain general for application to other Sun-like and low-mass stars.



\acknowledgments
Thanks for helpful discussions and technical advice from Marc DeRosa, Todd Hoeksema and Aline Vidotto. 
We thank Victor R\'eville for supplying the addition data from his MHD simulations.
This project has received funding from the European Research Council (ERC) under the European Union’s Horizon 2020 research and innovation programme (grant agreement No 682393 AWESoMeStars).
The sunspot number used in this work are from WDC-SILSO, Royal Observatory of Belgium, Brussels. 
Data supplied courtesy of the SDO/HMI and SDO/AIA consortia. SDO is the first mission to be launched for NASA's Living With a Star (LWS) Program.
Data provided by the SOHO/MDI consortium. SOHO is a project of international cooperation between ESA and NASA.
We thank the ACE MAG and SWEPAM instrument teams and the ACE Science Center for providing the ACE data.
We thank the \textit{Ulysses} FGM/VHM and SWOOPS instrument teams and the ESAC Science Data Center for providing the \textit{Ulysses} data.
We thank the developers of SHTOOLS, M. A. Wieczorek, M. Meschede, I. Oshchepkov, E. Sales de Andrade, and heroxbd (2016). SHTOOLS: Version 4.0. 
Figures within this work are produced using the python package matplotlib \citep{hunter2007matplotlib}.





\appendix
\section{A - Spherical Harmonic Decomposition}

\begin{figure*}
   \centering
    \includegraphics[width=\textwidth]{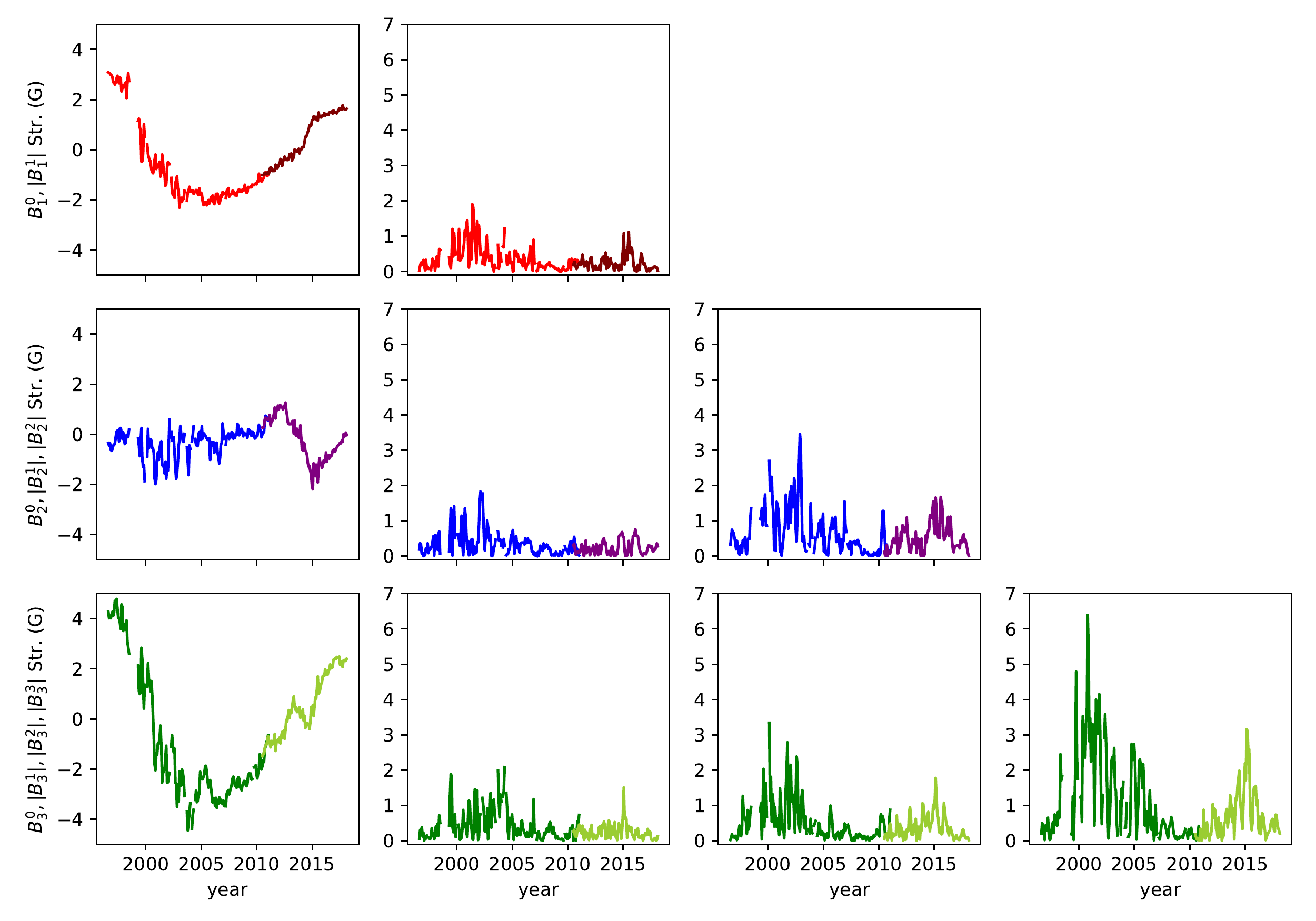}
     \caption{The evolution of the lowest order ($l\leq3$) spherical harmonic coefficients from SOHO/MDI and SDO/HMI. Rows display different spherical harmonic order $l$, increasing towards the bottom, and columns show increasing spherical harmonic degree $m$, increasing from left to right. The components for each $l$ and combined in quadrature to produce a single vale for each spherical harmonic, dipole, quadrupole, and octupole, shown in the top left panels of Figure \ref{solar_figure}. If only the axisymmetric components are used in our torque calculation, the predicted torque from equation (\ref{torque}) will have artificially small minima during polarity reversals. }
     \label{Coeffs_figure}
  \end{figure*}

For the synoptic magnetograms used in this work, we use the pySHTOOLS code \citep{wieczorek2011shtools} to deconstruct the observed surface magnetic field into its constituent spherical harmonic components, described by equation (\ref{spherical_harm}). This produces coefficients $B_m^l$ which weight each spherical harmonic mode $Y_m^l$ given a magnetic order, $l>0$, and degree, $-l\leq m\leq l$. Figure \ref{Coeffs_figure} displays the full spherical harmonic decomposition of both SOHO/MDI (1996-2010) and SDO/HMI (2010-2018) synoptic magnetograms. The strength of the axisymmetric dipole, quadrupole and octupole are given by $B_0^1$, $B_0^2$, and $B_0^3$ respectively. The absolute magnitude of the non-axisymmetric components are given, e.g. $|B_1^1|$, which incorporates both $m=1$ and $m=-1$ degrees. A deeper analysis of the variation in spherical harmonic components with time is available in \cite{derosa2012solar}.

\section{B - Validity of the Torque Predicted from Observed Stellar Rotation Rates}

The prediction of equation (\ref{eq_tausunrot}) is completely independent of any knowledge of solar magnetism or wind properties, or indeed even of the angular momentum loss mechanism itself, beyond the assumption of equation (\ref{eq_taurot}).  Because this is a robust and independent estimate of the solar torque, it is worth discussing the uncertainties that are inherent in this calculation.

First, the functional form of equation (\ref{eq_taurot}) should be taken as approximate. However, it is predicted by the stellar wind torque equations (\ref{torque})--(\ref{DQO_law}), if the stellar field strengths and mass loss rates depend on rotation rate as a power-law. This form is the usual assumption made in spin evolution models \citep[e.g.][]{kawaler1988angular, gallet2013improved, matt2015mass, amard2016rotating}.

Second, equation (\ref{eq_omegaconverged}) is an asymptotic solution for the converged spin rates.  A more precise calculation depends on the initial spin rate (which is unknown for the Sun), but the calculation can still be done by using the observed range of spin rates of young clusters.  For example, \cite{gallet2015improved} showed (using observations of \cite{agueros2011factory} and \cite{delorme2011stellar}) that, in the nearly 600 Myr-old clusters Praesepe and Hyades, 25\% of solar-mass stars rotate faster than 2.4$\Omega_\odot$ and 90\% percent rotate slower than 2.9$\Omega_\odot$.  This range of rotation rates at that age predicts a present-day solar torque in the range 5.9--6.3$\times 10^{30}$ erg.  Even extending further into the tails of the distributions of spin rates in those clusters implies a possible spin rate from 2 to 10 times $\Omega_\odot$, which gives a range of the present-day torque of 5.3--7.0$\times 10^{30}$ erg.

Third, the analysis assumes a constant moment of inertia and solid-body rotation. The assumption of constant moment of inertia is correct to better than 2\%, for solar-mass stars in the age range from 600 Myr to that of the Sun \citep{baraffe2015new}. The Sun and Sun-like stars are known to posses surface latitudinal differential rotation with an amplitude of approximately 20\% \citep{messina2003magnetic, barnes2005dependence, croll2006differential, matt2011convection}. The effects of latitudinal differential rotation should thus have a comparably small effect on the observed (single-value) surface rotation rates as being representative of the whole surface of the stars.  Helioseismic observations constrain the internal differential rotation profile also to an amplitude of approximately 20\% \citep{schou1998helioseismic, charbonneau1999stability}.  There is some evidence that the inner-most regions (10\% of the radius) of the Sun may rotate substantially faster than the surface \citep{garcia2007tracking, fossat2017asymptotic}, but this would only affect the total angular momentum by a small amount, compared to that inferred by assuming a solid-body rotation at the surface rate.  We have much less information about the internal rotation profile of Sun-like and younger stars.  It is possible that young stars' inner radiative zones rotate much more rapidly than the surfaces.  Rotational evolution models of \cite{gallet2015improved} predict this differential rotation can be substantial at ages of $\sim$100 Myr but decrease rapidly with time and has an amplitude of approximately 20\% by an age of 1 Gyr.  Thus it seems unlikely that differential rotation would affect the torque prediction by more than a few percent. However, given the uncertainty in internal rotation and angular momentum transport, it is worth noting that even in the most extreme case where the convective envelope is completely decoupled from the radiation zone implies a lower limit of 7.0$\times 10^{29}$ erg (calculated by putting the moment of inertia of the convective zone from \cite{baraffe2015new} into equation (\ref{eq_tausunrot})). 

Fourth, the average value of $p$ is fairly tightly constrained by the behaviour of observed spin rates over long timescales \citep{skumanich1972time, karoff2013sounding, barnes2016some, metcalfe2016stellar}. However, it is possible that the spin-down in time does not follow a single power law at all times \citep[e.g.][]{lanzafame2015rotational}, and there are relatively few observational constraints for stars with known ages between about a Gyr and solar age \citep{meibom2011kepler, meibom2015spin, barnes2016rotation, barnes2016some}. For example, the stellar wind torque model of \cite{gallet2013improved, gallet2015improved} had an asymptotic (i.e., late-time) value of $p\approx 3.2$, and they were able to fit available data.  \cite{van2016weakened} suggested that stars become abruptly less efficient at spinning down at around the solar age, which within the present formalism could imply larger values of $p$ (or generally speaking that the current torque could be significantly lower than predicted by equation (\ref{eq_tausunrot})).  It is not possible with the present analysis to rule out that the solar torque has undergone a recent transition, which would invalidate this calculation of the torque based on observed rotation rates of younger stars.

Despite the caveats listed above, equation (\ref{eq_tausunrot}) remains a robust estimate of the solar angular momentum loss rate, derived empirically from the observed rotation rates of other Sun-like stars. \\




\bibliographystyle{yahapj}
\bibliography{CyclesPaper}




\end{document}